# Order parameter of MgB$_2$: a fully gapped superconductor

H. D. Yang[1], J.-Y. Lin[2], H. H. Li[1], F. H. Hsu[1], C. J. Liu[3], S.-C. Li[4], R.-C[4]. Yu and C.-Q. Jin[4]

[1]*Department of Physics, National Sun Yat-Sen University, Kaohsiung 804, Taiwan ROC*

[2]*Institute of Physics, National Chiao Tung University, Hsinchu 300, Taiwan ROC*

[3]*Department of Physics, National Changhua University of Education, Changhua 500, Taiwan ROC*

[4]*Institute of Physics, Center for Condensed Matter Physics and Beijing High Pressure Research center, Chinese Academy of Science, P. O. Box 603, Beijing 100080, PRC*

We have measured the low-temperature specific heat $C(T)$ for polycrystalline MgB$_2$ prepared by high pressure synthesis. $C(T)$ below 10 K vanishes exponentially, which unambiguously indicates a fully opened superconducting energy gap. However, this gap is found to be too small to account for $T_c$ of MgB$_2$. Together with the small specific heat jump $\Delta C/\gamma_n T_c = 1.09$, scenarios like anisotropic $s$-wave or multi-component order parameter are called for. The magnetic field dependence of $\gamma(H)$ is neither linear for a fully gapped $s$-wave superconductor nor H$^{1/2}$ for nodal order parameter. It seems that this intriguing behavior of $\gamma(H)$ is associated with the intrinsic electronic properties other than flux pinning.

The recently discovered superconductivity in MgB$_2$ has stimulated new excitement in condensed matter physics [1]. MgB$_2$ possesses the highest $T_c \approx 39$ K among intermetallic compounds, and the strong isotope effect on $T_c$ clearly suggests an electron-phonon interaction [2]. Therefore, it is thought that $T_c$ of MgB$_2$ probably represents the upper limit of the phonon-mediated superconductivity [3]. However, to reach any conclusion, one requires better understanding of the superconducting properties of MgB$_2$. Up to now, there are several fundamental questions to be answered. For example, is MgB$_2$ a fully gapped superconductor? If it is, is it a conventional $s$-wave superconductor? In principle, the tunneling experiments can answer part of these questions. Nevertheless, the existing tunneling data are not consistent with each other, partly due to junction interface problems. On the contrary, the specific heat ($C$) measurements probe the bulk nature of the designated properties. If an exponentially vanishing electronic specific heat in the superconducting state $C_{es}$ is observed, a full superconducting gap $\Delta$ can be unequivocally established. Furthermore, the magnetic field $H$ dependence of $C_{es}$ would provide the information about the quasiparticle excitation associated with the vortex state, since MgB$_2$ is a type-II superconductor. Therefore, specific heat has been a powerful tool to investigate superconductivity among other methods. In the case of MgB$_2$, however, the present existing specific heat studies are not quite consistent with each other both qualitatively and quantitatively. To shed light on this issue, we have carefully conducted the low-temperature specific heat measurements on high quality polycrystalline $MgB_2$ prepared by high pressure synthesis. *The data clearly manifest a superconducting gap with no node at H=0. However, the same data show equally clearly that $MgB_2$ may not be a simple s-wave conventional superconductor.*

The polycrystalline $MgB_2$ sample was prepared by high pressure synthesis with pyrophyllite as the pressure transmitting medium at 3.0 GPa and 1000°C for 15 min. Details of the synthesis were described elsewhere [4]. The x-ray diffraction pattern showed a nearly single phase of $MgB_2$. In spite of the tiny amount of other phases like MgO and $B_2O$, the overall phase purity is better than that of the commercial alpha powder when the comparison was done. The superconductivity was first verified by the SQUID magnetization and the resistivity measurements. Both found the superconductivity onset around 39 K and the transition width less than 1 K. The resistivity at $T$=40 K is about 1 $\mu\Omega$ cm. The low resistivity, sharp transition, and the large specific jump $\Delta C$ at $T_c$ all indicate the high quality of the sample. $C(T)$ was measured from 0.6 to 50 K in the magnetic fields up to 8 T, with a $^3$He thermal relaxation calorimeter in which in which $C=(\Delta P/\Delta T)\tau$, where $\Delta T$ is the temperature change associated with the change of the heat power $\Delta P$ and $\tau$ is the thermal relaxation time constant. The precision of the measurement in the temperature range is about 1% through the whole temperature range. However, the electronic specific heat is better determined at low temperatures since at high temperatures a large phonon specific heat has to be subtracted. Details of the measurements and the system calibrations can be found in [5-7].

Fig. 1(a) shows the raw data of $C(T)$ of $MgB_2$ both at $H$=0 and 8 T. A clear anomaly around 39 K demonstrates bulk superconductivity in the sample. Better information of the anomalous specific heat jump at $T_c$ can be deduced from the deference between the superconducting state and the normal state specific heat $\Delta C=C_{es}-C_{en}$. Since $T_c\approx$20 K at $H$=8 T [8], $\Delta C$ above 20 K can be obtained by $C(H$=0)-$C(H$=8 T) . (The polynomial expression was used for $C(H$=8 T) during the subtraction.) The resulting zero field $\Delta C/T$ is shown in Fig. 1(b). The dimensionless specific jump $\Delta C/\gamma_n T_c$=1.09 at $H$=0, where $T_c$=39 K and $\gamma_n$=2.7 mJ/mol K$^2$ [9], is only 15% smaller than that in [10] and significantly larger than most of those in [2,9,11,12]. However, it is still small compared to the weak limit BCS value 1.43. This result may imply that Mg is not a simple strong-coupling BCS superconductor as one naively expects from its high $T_c$. Moreover, $\Delta C$ becomes zero at about $T_c/2$ as suggested in Fig. 1(b). A similar crossing temperature was also observed in [9,10]. For an isotropic s-wave superconductor with the coupling constant $\lambda$>1, the crossing temperature should have happened above 0.6 $T_c$ [13]. A curious dip of $\Delta C/T$ to a small negative value just above $T_c$ was observed. Interestingly, a similar dip with almost the same magnitude was also shown in [10]. This dip could be due to fluctuations and may deserve further investigation. Evolution of $\Delta C/T$ in the magnetic fields is shown in Fig. 2. It is seen that the bulk superconductivity continues to exist in magnetic fields up to at least 6 T, though only $\Delta C(T, H)$ up to 2 T is presented in Fig. 2 for clarity.

The low temperature data below 10 K reveal even more intriguing pictures. $C/T$ vs. $T^2$ below 10 K for $H$=0 to 8 T is shown in Fig. 3. For zero field data at very low temperatures, a remnant can not be excluded, probably due to small magnetic contribution or a few percents of normal state portion. Other than that, $C$ increases very fast with $T$. In magnetic fields, $C/T$ vs. $T$ is nearly a straight line at high temperatures. With increasing $H$, the intersection $\chi$ increases and the slope remains the same. All these results are expected, assuming an increase in the quasiparticle contribution. A more quantitative description for the data in magnetic fields can be given by $C(T, H)=\chi(H)T+C_{\text{lattice}}(T)+nC_{\text{Schottky}}(g\sim H/k_BT)$, where the second term is a 2-level Schottky anomaly $C_{\text{Schottky}}(x)=x^2e^x/(1+e^x)^2$. $C_{\text{lattice}}(T)=sT^3+uT^5$ represents the phonon contribution. The $T^5$ term due to the phonon dispersion is usually negligible at low temperatures, though inclusion of this term sometimes improves the quality of fit. The fits were carried out for data below 10 K at each $H$ by the above formula. Phenomenologically, the satisfactory descriptions for data were obtained, as exemplified by the solid lines in Fig. 3. By this method, $\chi(H)$, $n$, and $C_{\text{lattice}}$ can be obtained. The corresponding Debye temperature is $\Theta_D$=910 K which is close to that reported in [9,10]. The fits yield very similar $C_{\text{lattice}}$ for each $H$ as expected. The value of $n$ does not vary with $H$ very much, either. The averaged $n$=7.2×10$^{-5}$ mol$^{-1}$, which is much smaller than that in [9], again indicates the good quality of the sample, and makes the determination of $\chi(H)$ more reliable. In principle, one would be concerned with the expression of $C_{\text{Schottky}}$ which is, strictly speaking, only suitable for spin-1/2 paramagnetic centers. The existence of a small amount of Fe ions in MgB$_2$ was proposed [9]. Not knowing which oxide Fe forms, the spin of Fe ions can not be exactly determined. However, due to the small value of $n$, the resultant $\chi(H)$ is not expected to change drastically if another spin is chosen.

The electronic specific heat in the superconducting state at $H$ is $C_{\text{es}}(T)=C(T,H)-C_{\text{lattice}}(T)$. $C_{\text{es}}(T)$ vs. $T_c/T$ for $H$=0 was plotted in Fig. 4. The results unambiguously show that $C_{\text{es}}(T)$ vanishes exponentially at low temperatures. In Fig. 4, no attempt was made to correct the remnant. The fit from 3 to 10 K reveals $C_{\text{es}}(T)\propto\exp(-0.38T_c/T)$ which is demonstrated by the solid straight line in Fig. 4. This behavior of $C_{\text{es}}(T)$ clearly manifests the existence of an energy gap with no node. An exponential $C_{\text{es}}(T)$ was also reported in [10]. Deviations at very low temperatures are presumably due to the remnant.

With an apparent gap at low temperatures, MgB$_2$ may still not so simple as an isotropic $s$-wave superconductor. For the expression of $C_{\text{es}}(T)\propto\exp(-aT_c/T)$, the weak-coupling BCS theory predicts that the value of $a$ should be between 1.44 and 1.62 for the reduced temperature ($T/T_c$) range investigated in this Letter [14]. Obviously, the observed value of $a$ is four times smaller. One of the main points of this Letter is to establish the fact that *indeed a fully opened gap was observed at low temperatures, but this gap is too small to account for the high $T_c$ of MgB$_2$*. It is possible to reconcile this difficulty, at least qualitatively, by the scenario of either the anisotropic or multi-component order parameter for MgB$_2$. In both scenarios, the large gap(s) would be responsible for the $T_c$, while the low-energy excitation would be most influenced by the small gap. These scenarios are consistent with the theoretical considerations [15,16] and the results from the photoemission spectroscopy [17]. Since the sample in this study is polycrystalline, the present data are not likely to distinguish either of these scenarios. It is now clear that the small $\Delta C/\gamma_n T_c=1.09$ is probably intrinsic in order to be in accord with the observation of a small gap at low temperatures. Previously, the small $\Delta C/\gamma_n T_c$ of Mo was considered as manifestation of gap anisotropy [18]. In the multi-gap (say, two-gap) scenario, both the large and small gap should open at $T_c$. If the small gap were BCS-like and opened at $T\approx 10$ K, $C_{es}$ should have deviated from the exponential behavior above 5 or 6 K. Furthermore, no corresponding anomaly was observed around this temperature both in this work and in [9], though a possible signature was suggested in [10].

The magnetic field dependence of $\gamma(H)$ is even more puzzling. For a gapped superconductor, $\gamma(H)$ is expected to be proportional to $H$ [19]. Apparently, $\gamma(H)$ is not linear in $H$ as seen in Fig. 5. For nodal superconductivity, $\gamma(H) \propto H^{1/2}$ is predicted [20], and was verified by several studies in cuprates [5,6,21-23]. Fitted to the power law, $\gamma(H) \propto H^{0.23}$ can be found as shown by the dashed line in Fig. 5. Therefore, this nonlinearity in $\gamma(H)$ may be due to origin(s) other than the nodes of the order parameter. Phenomenologically, the small power is caused by the rapid increase in $\gamma(H)$ at low fields, and is likely to related to the small gap mentioned in the previous paragraph. This possibility is suggested by the poor description of high field data in Fig. 5 by the power law. Since this curious behavior was observed in hot-pressed polycrystalline sample here and the powder sample in [10] or the sintered one from the commercial powder in [9], it is not likely to originate from pinning mechanism. More likely, in addition to the possible cause of a small gap, it is associated with the quasiparticle states in the cores or the unusual magnetoresistance reported in [8]. Extrapolating $\gamma(H)$ by this power law fit, $\gamma(H)=2.6$ mJ/mol K$^2$ at $H=14$ T, which is consistent with $\gamma_n$ in [9]. Other than in MgB$_2$, it is noted that a nonlinear $\gamma(H)$ in $H$ was also observed in the $s$-wave superconductors like NbSe$_2$ [24]. Therefore, the thermodynamics of the mixed state in the gapped superconductors apparently is not well understood and might provoke new interest on this issue.

To conclude, we have presented convincing evidence for a fully opened gap in MgB$_2$. The low temperature $C_{es}$ in this work is qualitatively similar to that in [9,10]. However, it is quantitatively consistent with an exponential behavior as reported in [10] rather than the $T^2$ dependence in [9]. The gap deduced from the exponential $T$ dependence is too small, together with the small $\Delta C/\gamma_n T_c$, to account for $T_c=39$ K in the isotropic $s$-wave model. Other scenarios like anisotropic or multi-component order parameter are called for to reconcile. Isotropic or not, the nonlinear $H$ dependence of $\gamma(H)$ can not be easily explained. The present study suggests that MgB$_2$ is not a simple conventional superconductor. To understand the origin of its high $T_c$, more work on the fundamental properties, especially on the complexity of order parameter is indispensable.

This work was supported by National

Science Council of Republic of China under contract Nos. NSC89-2112-M-110-043 and NSC89-2112-M-009-052. We would like to thank Y. Y. Hsu and H. C. Ku for magnetization measurements and C. Y. Mou for discussions.

## Figure Captions

Fig. 1. (a) $C(T)$ of $MgB_2$ both at $H$=0 and 8 T. The anomaly around 39 K manifests the bulk superconductivity. $C/T$ vs. $T$ was plotted in the inset to show the thermodynamic consistency. (b) $\Delta C(=C(H=0)-C(H=8\ T))/T$ vs. $T$. The dashed lines are determined by the conservation of entropy around the anomaly and used to estimate $\Delta C/T_c$. The dot line at low temperatures is an exponential extrapolation of data between 3 and 5 K. Inset: entropy difference $\Delta S$ by integration of $\Delta C/T$ according to the data or the extended dot line.

Fig. 2. Evolution of $\Delta C = C(H)-C(H=8\text{ T})$ in magnetic fields.

Fig. 3. $C(T,H)/T$ vs. $T^2$ below 10 K. The solid lines demonstrate the quality of the fit.

Fig. 4. $C_{es}=C-C_{lattice}$ of MgB$_2$ in the superconducting state is plotted on a logarithmic scale vs. $T_c/T$. The exponential dependence on $1/T$ is evident. The straight line is the fit from 3 to 10 K.

Fig 5. The magnetic field dependence of $\chi(H)$. The solid line is a fit to the power law.

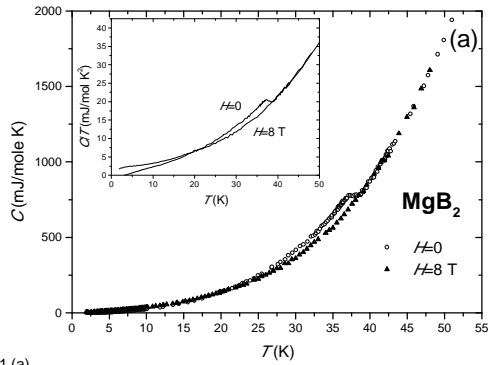
Fig. 1 (a)

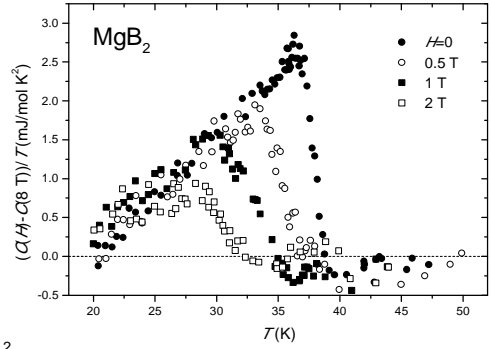
Fig. 2

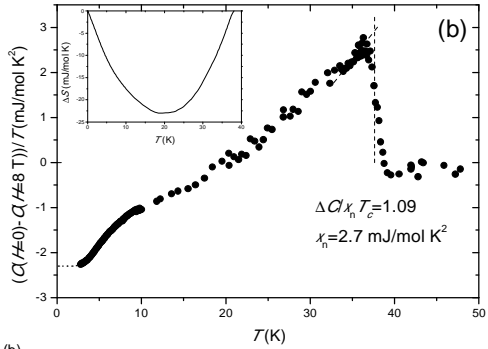
Fig. 1(b)

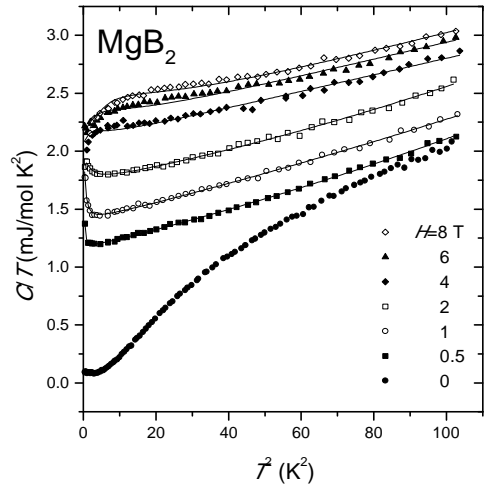

Fig. 3  Yang et al.

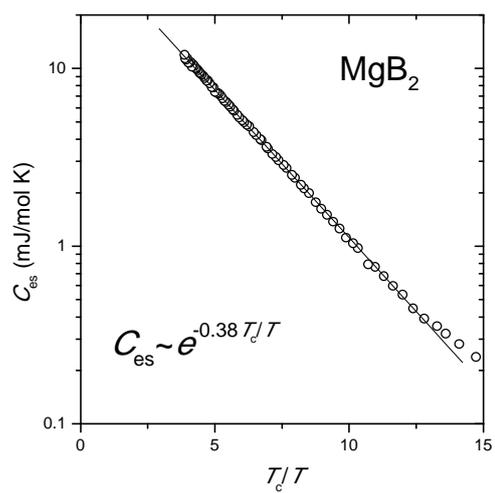

Fig. 4    Yang et al.

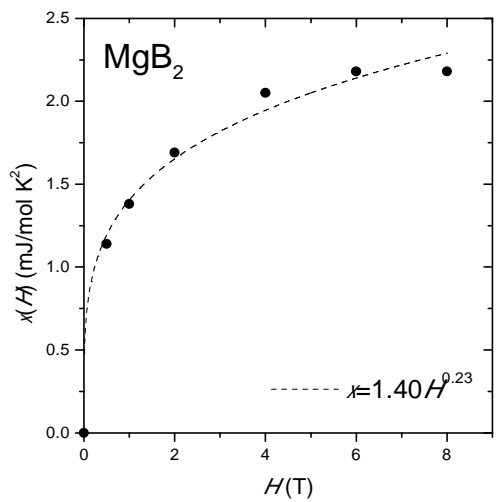

Fig. 5
Yang et al.